\documentclass[a4paper,openacc]{rstransa}
\usepackage{amsmath,amssymb,amsfonts,amsthm,graphicx,endfloat,endnotes,setspace}
\usepackage{verbatim,geometry,times,helvet,courier,bm,url,babel,dcolumn}
\usepackage{aas_macros}
\usepackage{atbegshi}
\AtBeginDocument{\AtBeginShipoutNext{\AtBeginShipoutDiscard}}
\jname{rsta}
\Journal{Phil. Trans. R. Soc}
\Doi{2020.0568}

\newcommand{\mydeg}{{$^{\circ}$}}

\begin{document}
\title{Occultation constraints on solar system formation models}
\author{Marc W. Buie$^{1}$, John M. Keller$^{2}$, David Nesvorn\'y$^{1}$, and Simon B. Porter$^{1}$}
\address{$^{1}$Southwest Research Institute, 1301 Walnut St., Suite 400,
Boulder, CO 80302, USA\\
$^{2}$University of Colorado, Boulder, USA}
\subject{Kuiper Belt, occultations, solar system formation}
\keywords{Arrokoth, Polymele, streaming instability}
\corres{Marc W. Buie\\
\email{buie@boulder.swri.edu}}

\begin{abstract}
The process by which a system of non-luminous bodies form around a star is fundamental to understanding the origins of our own solar system and how it fits into the context of other systems we have begun to study around other stars.  Some basics of solar system formation have emerged to describe the process by which dust and gas around a newly formed star evolve into what we see today.  The combination of occultation observations and the flyby observations by New Horizons of the Cold-Classical Kuiper Belt Object (CCKBO), (498958) Arrokoth, has provided essential new constraints on formation models through its three-dimensional shape.  We present a case that an occultation-driven survey of CCKBOs would provide fundamental new insight into solar system formation processes by measuring population-wide distributions of shape, binarity, and spin-pole orientation as a function of size in this primordial and undisturbed reservoir.
\end{abstract}


\begin{fmtext}
© The Authors. Published by the Royal Society under the terms of the
Creative Commons Attribution License http://creativecommons.org/licenses/
by/4.0/, which permits unrestricted use, provided the original author and
source are credited.
\end{fmtext}
\maketitle

\section{Introduction}

Occultations have a very long observational history.  Probably the most spectacular and familiar example is a total solar eclipse, with documented observations going back thousands of years.  Through this opportunity, fundamental research on our Sun was possible long before telescopic and space-based observatories began.

Lunar occultations of stars were used for finding binary stars and measuring stellar diameters as well as characterizing topography on the limb of the moon.  Here, high time resolution occultation data translated into high spatial resolution data that was otherwise unattainable.

Occultations of stars by other more distant solar system objects were obvious phenomena worthy of consideration.  Successful prediction and observation of these occultations require high-quality information about stellar positions, orbits of the occulting bodies, and sufficiently capable instrumentation.

The earliest successful visual occultations were by large objects, e.g. Juno (1958 Feb 19) and Pallas (1961 Oct 2) \cite{Taylor1962}.  Even though events like these could be predicted and observed, opportunities were rare because of the low precision stellar positional catalogs and inaccuracies in orbit estimates.  Later, the giant planets were studied due to the atmospheric signatures from occultations.

Extending the use of occultations to the small body population has been a goal since the middle of the 20th century.  We will focus on this application of occultations in this paper.  A more complete overview of all aspects and history of occultations can be found in reference \cite{Sicardy2024}.

With the development of photoelectric photometers, it was possible to record data fast enough to take full advantage of an occultation.  Over the years, the instrumentation has become more capable and easier to use, all at a vastly lower cost.  With lower cost comes greater accessibility, and no longer requires a professional commitment and level of resources to be involved.

The remaining challenge to the widespread use of occultations to study small bodies was in having a sufficiently precise catalog of stellar positions.  Stellar catalogs consistently improved over the years but limited the success rate while also requiring significant effort to overcome their limitations.

In 2013, the Gaia space observatory was launched and embarked on a multi-year project to build the best stellar catalog ever attempted.  The goal of the mission was to build a kinematic map of our galaxy by measuring the positions, sky-plane velocities, and distances through stellar parallax.  Gaia measurement requirements included having the sensitivity to measure these properties even on the far side of the galaxy.  The result was a catalog nearly three orders of magnitude more precise than any prior work.  For occultations, it is equally important that the positions as a function of time are also provided.  On the scale needed for occultation predictions, the stellar motions are essential.

In 2018, the second data release (DR2) from the Gaia mission provided the stellar catalog needed for occultation predictions\cite{Gaia2018} with high-quality proper motions.  Armed with this amazing data product comes an enormous opportunity to investigate the small body population in our solar system.

This paper will describe the essential basics of small-body occultation observations that are now possible.  We will show some current open questions, specifically in the realm of solar system formation models, that can be addressed with a large and coordinated occultation survey.  We will also present a discussion of the challenges remaining in trying to collect the data that are now within reach for the very first time.

\section{Guiding Principles of Occultations}

Occultation observations are deceptively simple: Measure the brightness of a star as a function of time.  For a positive detection, the signal will drop below what the un-occulted star presents.  With a negative detection, no change in the star signal is seen.  For solid body events, the star is completely obscured from view for a short time.  For objects with atmospheres, the star dims at a rate that can constrain atmospheric properties.  This discussion will focus on solid body events since the class of objects of interest to this discussion are all too small and cold to have atmospheres.

Successful occultation observations are driven by a few important factors.  First, the area to be covered to reveal the shape of the body is the same size as the object.  This shadow is slightly distorted when projected onto the Earth, but its size, when mapped back into the sky reference frame, is the exact profile of the object.

The second factor is the star, both its size and position.  Uncertainties in stellar positions prior to the results of the Gaia mission posed a severe challenge and limited the number of successful occultations.  For most occultation predictions now, the positional error of the target star can all but be ignored.  The closest stars, which also usually appear to be bright, can be resolved by an occultation.  If the star is resolved, the usually sharp drop in brightness will be more gradual and this effect must be calibrated and removed to return the size and shape information.  Information from the Gaia catalog can permit the estimation of the expected size.  A remaining complication occurs when the star is an unrecognized double.  In this case there will be two (or more) distinct shadows whose location differs from the nominal prediction.  We will not discuss this case further since it is a minority outcome and the strategies for avoiding or using such opportunities are complicated and beyond the scope of this work.  Note that we are not concerned with the finite spatial resolution limit set by Fresnel diffraction.  For the sizes of objects we will discuss, this factor can largely be ignored.

The third factor is having an accurate orbit for the asteroid.  Not all the known or even numbered objects have sufficiently precise orbits.  A well-observed object that has an observational record at least 10 years long is usually a good candidate for consideration, but there can be special cases with shorter arcs.  Regardless of the observational record, it is important to successful occultation work to have a firm understanding of the positional uncertainty at the time of a predicted occultation given an orbit estimate.

The importance of the positional accuracy depends on the distance to the object.  A fixed spatial distance subtends an ever smaller angle as the distance increases.  Most well observed objects in the main asteroid belt can be predicted well and the uncertainty in the object's size is a more important factor.  Positional uncertainties based on traditional astrometric measurements begin to dominate at 5 AU for the Jupiter Trojans and predictions become progressively more uncertain out into the Kuiper Belt.

For example, an object with a diameter of 100~km at a distance of 2.5~AU will subtend 55 mas.  Positional uncertainties are comparable to this angular size, smaller for low-numbered asteroids with very long observational histories and slightly larger for shorter-arc orbital estimates.  Understanding the implications of these numbers is easier when converted to an observing strategy.  For this example, consider a goal of getting 4 chords evenly spaced across the object.  A station at exactly 50 km from the shadow center will have an undetectable interaction with the target star.  To compensate and not count ultra-short chords, reduce the diameter to an effective coverage footprint that is 80\% of the true diameter when computing the spacing.  The spacing ($\delta$) needed will thus be $\delta = 0.8*D/c$, where $D$ is the diameter of the asteroid, and $c$ is the number of evenly spaced chords desired.  For this example, $\delta=20$~km.

The second key factor in the deployment strategy is the spread of stations.  This choice in this case is driven by the range of uncertainty ($\sigma$) to be covered.  The number of stations to deploy is thus $N = 1 + (D + 2n\sigma) / \delta$ where $n$ is the coverage desired in units of $\sigma$.  Assume for this case that $\sigma=100$~km (same as $D$ and 55~mas in this example).  A 3-$\sigma$ deployment would have a spread of 700 km and require 36 stations with $\delta=20$~km.  Obviously, in this case, a successful outcome depends strongly on how much of the probability envelope is covered.  A 1-$\sigma$ deployment would only require 16 stations, but the probability of getting $c=4$ chords drops from 98\% to 68\%.  The choice of $c$, $n$, and $\delta$ must be balanced against the cost and effort of the number of stations to be deployed versus the information on the limb profile desired.

The challenge becomes clear when scaling this case up to the Kuiper Belt.  For the prior example, change the distance to the asteroid to be $\Delta=43$~AU\null.  Now the same uncertainty gives $\sigma=1720$~km. The 3-$\sigma$ deployment now has a spread of 10,420~km and would require 522 stations.  Fortunately, many KBOs have uncertainties smaller than our example and several successful events have been recorded.  Reaching a success rate at an interesting level for a survey of outer solar system objects will require a focused and strategic effort.

This example is harder than those cases pursued so far.  A more realistic case for current methods is starting from a factor of three lower uncertainty.  Continuing this case, we will take the baseline uncertainty at the start of 20~mas, corresponding to $\sigma=625$~km at 43~AU\null.  This level rescales a 3-$\sigma$ case to a $\sim$3850~km spread and 194 stations -- still a formidable number of stations.  Dropping the number of chords to $c=2$ would require 97 stations.

One strategy for collecting such multi-chord observations of KBOs takes an incremental approach built on multiple occultation observations.  The first campaign will have a relatively large uncertainty and would normally require an impractical number of stations for a high-resolution profile as shown in the examples above.  A more practical plan in this case is to change the goal to get just one chord.  This is the scenario that the RECON project \cite{Buie2016} was designed for with a spacing of 50 km and a success probability of 30\% that was targeting 100 km objects.   A successful event provides an astrometric position that is no worse than the angular size of the object.  For KBOs, this measurement provides a very strong constraint on the orbit and will lead to a significant improvement in subsequent occultation predictions.

A second campaign takes advantage of the orbit update from the first.  Now, the prediction uncertainty will typically be closer to the size of the object. Taking this case, the new spread for a 3-$\sigma$ deployment will drop to $\sim$700~km, comparable to the main-belt case first discussed.  With this improvement, 30 stations with a 24~km spacing will now return a 3-chord result.  This result will establish important size and shape constraints and another strong astrometric constraint for the orbit.  After two events, the uncertainty of the position will be limited by the Gaia catalog uncertainty for the star (around 0.1 mas).

The third campaign can deliver a detailed shape with a much tighter spacing with far fewer deployed stations.  It is not unreasonable to expect the uncertainty to drop by a factor of ten.  Now a 30-station deployment can cover 3~$\sigma$ with a 5.5 km spacing and expect to get 15 positive chords across the central 80\% of the object.

Based on experience with dense observations of Lucy mission targets, this level of spatial information on the shape will easily discern between spherical bodies and more complex shapes such as prolate ellipsoids, oblate ellipsoids, contact binaries, or other more complex and highly distorted shapes.

These examples present a necessarily idealized discussion.  If one had control over the timing of the occultations, they would be scheduled over a short time span since the positional knowledge degrades with time between occultation campaigns.  However, the cases shown are representative as long as the time between occultations is short compared to the length of the time over which prior astrometry spans.  For an object with a 20-year arc, occultations spaced apart by a year will work pretty well.  In specific cases, the details will, of course, matter.

A second and inevitable result of such a set of occultation observations is constraining unseen mass.  The spatial resolution of even the first campaign can reveal close equal-mass binaries.  In fact, even if the uncertainty were smaller for the first campaign, it would be strategically important to cover a large area with a coarse array just for this search.  Even if no extra objects are seen, having the desired set of three occultation measurements will provide astrometry that can set strong limits on unseen objects and suggest objects worthy of additional followup where the orbit fit residuals are unusually high.

The occultation method applied to KBOs has an additional advantage over lightcurves.  Constraining the three-dimensional shape and the orientation of the spin pole from lightcurves requires a broad range of viewing geometries.  These geometries are possible for Jupiter Trojans inward, but more distant objects have very long orbital periods, making this type of inversion impractical without multi-generational investigations.  Occultation data is much more sensitive to oblate shapes with a limited data set.  A full three-dimensional shape is still difficult, but statistically based determinations of the population are far stronger when including occultation constraints. Occultation observations can also provide irregular or contact binary shapes where no measurable lightcurve exists due to looking down the rotational pole \cite{Showalter2021}.

Some general attributes of occultation studies of KBOs are worth summarizing to illustrate the level of support required.  The three-campaign model for the characterization of a single KBO is a useful starting point.  The supplied examples refer to a 30-station deployment and for the sake of this discussion treat this as a baseline campaign size.   Given a two-person team per telescope, a single campaign would involved 60 participants.  Getting the three campaigns on a single object would then bring the total number of participants to 180.

The cost of deployment comes down largely to the cost of supporting the participants.  The support model developed for New Horizons and Lucy occultation campaigns brings people together for training, practice, and eventual deployment.  This model includes four nights and can bring a participant from no prior knowledge to a successful observation at the end.  The costs of this approach are driven entirely by the cost of supporting the participants while on the campaign, usually lodging, meals and transportation.  All equipment needed is provided by the project.  So far, there are no apparent limits in the ability to recruit as many participants as needed.

An alternate style of campaign is to involved those already in possession of suitable equipment and experience, such as IOTA members.  The advantage here is that the costs can be very low, often zero.  However, such low-cost campaigns are not easily scaled to a large project.  If the active IOTA community were 10-100 times bigger, the situation would be different.

Describing a full cost model is beyond the scope of this work.  However, an example case will serve to motivate the discussion.  A 30-station, 60-participant campaign would cost \$108k.  (All costs are given in US dollars appropriate for 2025.)  This is for a four-night campaign with half of the participants flying into the region and an average of 800~km driving for each team and is relevant for nearly anywhere in the world appropriate for observing occultations.  Given this canonical cost, a single object would cost \$324k.  Shipping costs for equipment are small compared to the participant costs.

A full survey design is a subject for future work but here we provide a fiducial case.  If our goal is to measure shapes and sizes of 1000 KBOs, the participant support cost would be \$324M.  Simple estimates that these costs would be 85\% of the total survey cost.  The rest of the costs would come from organization and planning of the campaigns for a grand total of \$381M.  Such a project would, in the end, involve 180,000 participants, not including the support staff.  Note that if run as a 10-year program, this would imply an annual operations cost of \$5.7M/year.

Such a large survey could arguably address just about any hoped for population constraint though such final results would require separate work to remove observational biases.  A project of this scale is not reasonable using either IOTA or the professional community as observers.  Involving the general public is the only pathway to a project of this scale.

\section{Arrokoth, Polymele, and Formation Models}

Planet formation is a fundamental scientific problem in astrophysics. We want to know how the Solar System planets formed, how planets populate the universe, and how they become habitable and might host extraterrestrial life. An important stage of planet formation is the formation of ``small planets'', or planetesimals, that are 1-1000 km in size. There are several populations of planetesimals in the Solar System: asteroids, Jupiter Trojans, Kuiper Belt Objects (KBOs), and the Oort cloud. The size distributions of asteroids, Jupiter Trojans and KBOs provide important constraints on planetesimal formation. In all cases there is a marked excess of 100-km class bodies, which is thought to reflect the preferred size of objects that are assembled by planetesimal formation processes \cite{Klahr2020}.

The information obtained from occultation observations offers an exciting possibility of testing planetesimal formation theories. In a classical hierarchical coagulation model (e.g., \cite{Kenyon2008}), planetesimals gradually grow by accretional collisions of smaller bodies to eventually reach sizes of tens to hundreds of km. The stochastic nature of this process is expected to produce a wide range of planetesimal shapes (think random Gaussian spheroids). Another class of formation theories invokes rapid formation of planetesimals by instabilities. For example, the streaming instability (SI) is a mechanism that aerodynamically concentrates small particles in a proto-planetary disk \cite{Youdin2005}. Planetesimals form in this model when the cloud of small particles gravitationally collapses \cite{Nesvorny2010}. This process is expected to be more orderly and lead to planetesimals with diagnostic characteristics, as we discuss below.  This quick summary is inadequate to capture the full scope of prior work and thinking but references \cite{Morbidelli2020,Gladman2021} would be an excellent source to learn more.

In published SI simulations, the particle clouds inherit rotation from the turbulence in the parent proto-planetary gas disk. As the angular momentum must be conserved during collapse, the planetesimals that emerge from this process are expected to be fast rotating, flattened or even elongated (to maximize their angular momentum contents \cite{Stern2023}). Often, the initial angular momentum of the particle cloud is too large for a single compact object to form; such an object would be rotationally unstable and split.  The SI model thus predicts the formation of planetesimal binaries for which a large share of the initial angular momentum can be deposited in the mutual orbit. The binaries -- again depending on the initial angular momentum of the particle cloud -- can be contact, tight or loosely bound. At least some predictions of the SI model neatly match the population characteristics of equal-size binaries observed in the Kuiper belt \cite{Grundy2019,Nesvorny2019}. 

The occultation observations of Arrokoth \cite{Buie2020}, followed up by direct imaging from the NASA New Horizons spacecraft \cite{Stern2019}, provided a spectacular test of planetesimal formation theories. Arrokoth is a member of a dynamically cold population of KBOs, known as Cold Classicals, at $\sim 45$ au from the Sun. Here, there is a lot of empty space, and planetesimals can survive over the age of the Solar System without suffering major reshaping impacts \cite{Greenstreet2019,McKinnon2020,Spencer2020}. In contrast, for example, the population of main belt asteroids at $\sim 2$-3.5 au have been extensively changed by impacts \cite{Bottke2015}. The shapes and other characteristics of Cold Classicals, Arrokoth included, must date back to their formation.  This conclusion is based on the implicit assumption that collisional processes tend to drive all starting shapes toward a somewhat spherical end state.  For this to be true, observations of main-belt asteroids will continue to show a dearth of oblate objects compared to KBOs.

Arrokoth is a delicately balanced contact binary. The larger lobe, Wenu, is strongly flattened (21.2 km $\times$ 19.90 km $\times$ 9.05 km \cite{Keane2022}) and shows conspicuous large mounds on the surface \cite{Stern2023}.  The smaller lobe, Weeyo, is more spherical (15.4 $\times$ 13.8 $\times$ 9.8 km\cite{Spencer2020}) and bears a large depression feature which is likely an impact crater. Alignment of the moments-of-inertia and spin poles within 10\mydeg\ for the two lobes suggests the separate formation of the two lobes on a bound orbit \cite{McKinnon2020}, physical alignment from aerodynamic drag and/or tides \cite{Lyra2021}, and slow, grazing merger ($<5$ m/s); \cite{Marohnic2021}. This most likely implies early formation of Arrokoth in the solar nebula \cite{McKinnon2020}, probably by the SI-triggered collapse of a particle cloud.  It would be extremely difficult to explain how Arrokoth's characteristics could arise from stochastic collisions invoked in the hierarchical coagulation model. 


Polymele, a member of the Jupiter Trojan population and one target of the NASA Lucy mission \cite{Levison2021}, also has a strongly flattened shape. Polymele was well observed during the occultation campaign in 2022 \cite{Buie2022} after two lower resolution range-finding precursor campaigns.  Shape models based solely on lightcurves predicted a nearly spherical body\cite{Levison2021}.  The occultation results require an oblate ellipsoidal shape with approximate dimensions of (27.0 $\times$ 24.4 $\times$ 10.4 km; \cite{Buie2022}), revealing that this object is even more flattened than the large lobe of Arrokoth.  As we discussed above, the SI-triggered gravitational collapse of rapidly rotating particle clouds is expected to give rise to a population of strongly flattened planetesimals \cite{Stern2023}. So, in this sense, Arrokoth and Polymele provide important clues about planet formation: they are likely a product of the same planetesimal formation process as deduced from the occultation-derived shapes. 

It is difficult to understand, however, how Polymele could survive without suffering at least some shape-changing impacts.  In fact, the occultation data shows Polymele has a more irregular shape than Arrokoth \cite{Buie2022}.  Unlike Arrokoth and Cold Classicals, Jupiter Trojans are presumed to have evolved through several stages when shape-changing collisions may have occurred.  The Lucy mission flyby of Polymele in 2027 will provide essential ``ground-truth'' of the geologic context and cratering history that can be connected to the occultation shapes \cite{Levison2021}, but we expect Trojans to be intermediate in impact modification between main-belt and CCKBO populations.  The survival of Polymele's oblate shape is already an indication of less collisional processing than the main-belt asteroids.  In situ investigations from Lucy will provide a detailed look at five objects.  Other means will be required to extend this small sample to the populations as a whole.

The birth location of Polymele is probably the outer planetesimal disk originally at $\sim20$-30 AU (the primordial Kuiper belt, \cite{Nesvorny2018}), from where they dynamically evolved to be eventually captured as Jupiter Trojans \cite{Emery2015}. The outer planetesimal disk is thought to have been relatively massive (15-20 Earth masses), and that is why this stage -- before the disk was dispersed by migrating Neptune -- is thought to be the most important for Polymele. If Polymele's shape survived this stage almost intact, that would be an important constraint on the lifetime of the outer disk. It would indicate that the disk was relatively quickly dispersed by Neptune (in less than 10 Myr?, short-lived disks imply fewer impacts).  This shows how occultation observations can provide useful constraints on the early evolution of the Solar System.    
    
Looking toward the future, the proposed systematic simulations of SI will reveal how planetesimal and planetesimal binary properties depend on proto-planetary disk conditions. These simulations will have high-enough resolution to determine, on a population level, the planetesimal shapes, spins, obliquities, occurrence of contact and separated binaries, etc. In specific cases, soft-sphere PKDGRAV modeling will be performed to reveal the formation of surface features, such as Arrokoth's mounds \cite{Stern2023}. 

Important size dependencies will be identified in this effort. Various size dependencies are expected from the size-dependent nature of physical processes involved in planetesimal formation. In at least some instances, these processes and their implications are understood. For example, the strength of (size-dependent) aerodynamic gas drag determines whether a binary orbit will shrink with the two binary components evolving into contact (thus forming a contact binary) \cite{McKinnon2020,Lyra2021}. This is probably the reason smaller planetesimals such as Arrokoth are contact binaries, but larger 100-km class KBOs are found in more loosely bound binaries, at least with current data.

In other instances, important size dependencies are suggested by occultation observations, but not understood in theory. For example, large 100-km class Jupiter Trojans and KBOs are considered to be relatively spherical.  In this $\sim10$-km class, Polymele and Wenu are strongly flattened\cite{Spencer2020,Buie2022}.  Not enough data exist to say if these cases relate to a general trend.  If so, this could mean that whether a planetesimal becomes more spherical or more flattened probably depends on the object's size, and therefore on the mass of the particle cloud that collapses to make that object. For example, the less massive clouds could rotate more vigorously and lead to small planetesimals with more flattened shapes (but this is something that has yet to be demonstrated).  However, the common assertion for larger objects should be tested as well.

Whereas the existing data for Arrokoth and Polymele provide important support of the SI-driven formation of planetesimals, it is difficult to reach firm conclusions from only a handful of KBOs and Jupiter Trojans that have been targeted in detailed occultation surveys. But imagine what could be done if high-quality occultation observations were conducted for many KBOs and Jupiter Trojans, resulting in reliable shape models for hundreds of them - what would it tell us? 

The impact of a population-wide occultation survey will be revolutionary. For example, the distribution of shape characteristics, ranging from the strongly flattened planetesimals such as Polymele and Arrokoth to more spherical bodies, would give us a handle on the distribution of angular momentum in parts of the turbulent proto-planetary disk where/when planetesimal formation happened. By comparing these characteristics with detailed SI simulations, it will be possible to constrain fundamental SI processes operating in the proto-planetary disk (e.g., level of turbulence; with only limited constraints at the present time, the SI parameters have to be blindly assumed in the existing SI simulations). Such fundamental constraints are not available from elsewhere. Whereas the currently available occultation data hint at various size dependencies of planetesimal formation processes, the next-level occultation surveys will characterize these dependencies. 

\begin{figure}
    \includegraphics[trim=60 40 10 20,width=\textwidth]{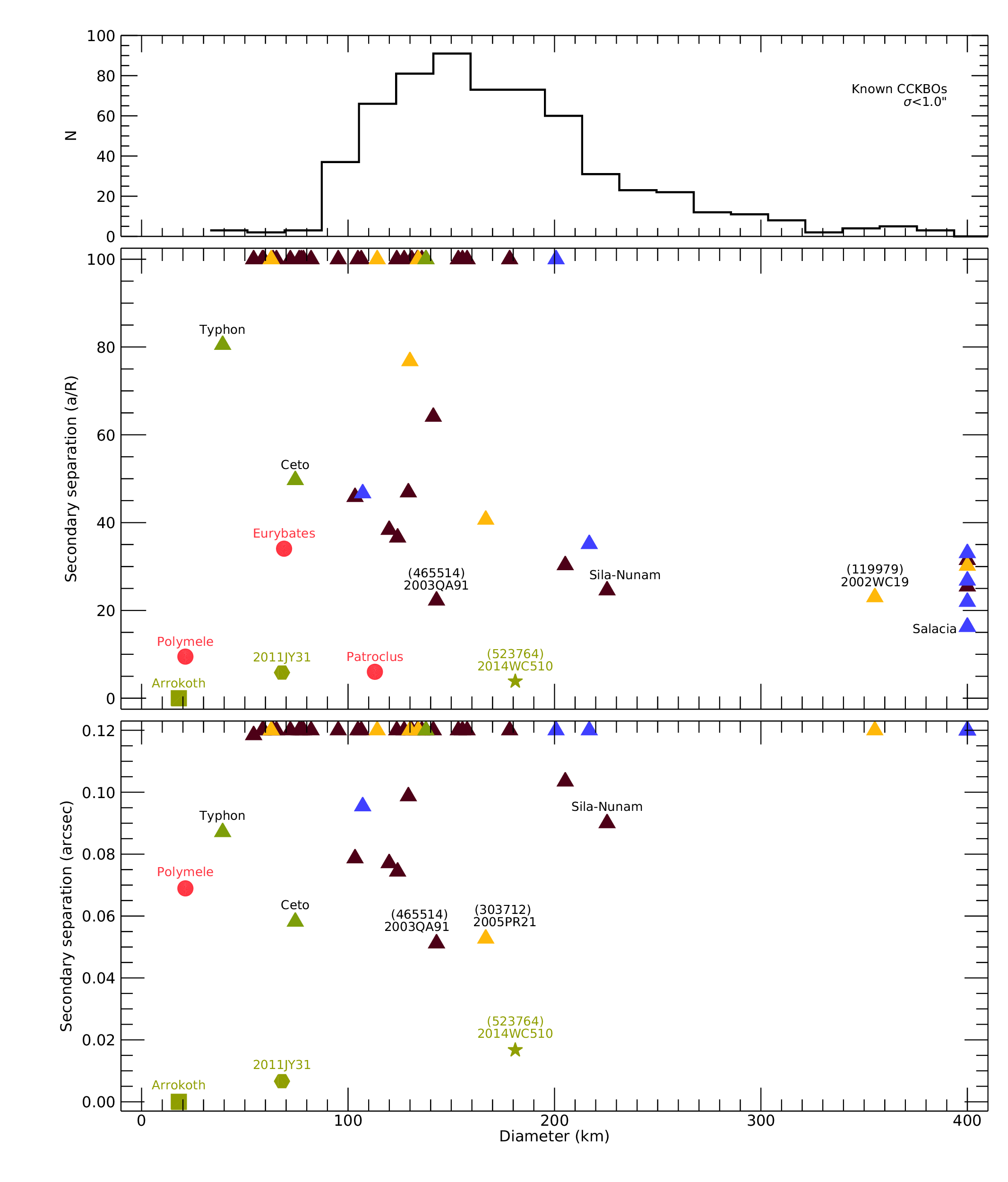}
    \centering
    \caption{Summary of KBO binaries and the context of Cold-Classical objects.  The top panel is a histogram of known Cold-Classical KBOs with current ephemeris uncertainties less than 1 arcsec.  The middle panel shows the dynamical properties of known KBO binaries shown with their orbital separation in units of primary radii plotted against the estimated diameter of the primary.  The bottom panel shows the observational properties of the same known objects.  Values clumped at the top and right plot beyond the range shown.  See text for a full description.}
    \label{fig:bstat}
\end{figure}

Occultation surveys would be a powerful tool for detection and characterization of contact and tightly bound binaries, including close-in satellites.  Currently known binaries are limited by the spatial resolution limit, even from space-based observatories, and with separations that are dynamically tight (few percent of a Hill radius) but still thousands of kilometers\cite{Noll2020}.  The tightest binaries in reference \cite{Noll2020} are (120347) Salacia and (174567) Varda, both with separation of $\sim$16R\null. There are claims that contact binaries can be more easily detected by lightcurve observations, but often the lightcurve inversion is not unique, or the results have only statistical validity \cite{Showalter2021}.  Occultation-based probes will cover the smallest end of the stability range \cite{Leiva2020}.

Figure~\ref{fig:bstat} shows a summary of known KBO binary systems plus a few other instructive examples.  The top panel shows a histogram of the known CCKBOs with positional uncertainties less than 1$''$ as of 2025 January 06.  This sample contains 617 objects.  The middle panel shows the known binaries in a dynammical context where more tightly bound objects are toward the bottom of the plot.   The bottom panel shows the known binaries in an observational context giving the maximum angular separation for the closest observing distance during 2025.   The scale of the bottom panel is adjusted to show no more than 3-pixel separation for the WFC3 camera on HST\null.  The triangles denote objects from reference \cite{Noll2020} and these were all discovered with direct imaging.  The total in this sample is 51 objects.  These triangles are color coded by their dynamical classification: CCKBOs are dark red (33), resonant objects are yellow (7), scattered and hot classicals are blue (8), and Centaurs are green (3), where the numbers indicate how many of each type is contained within the sample.  Six additional objects are shown: three Lucy targets (red circles), binary discovered by New Horizons (green hexagon), resonant KBO binary discovered by occultation, and contact binary discovered by occultation (green square).  The tightest binaries from the KBO sample are labeled across the sampled range of sizes.

The only known contact binary in the outer solar system, Arrokoth, is shown as the green square.  This is the smallest object in the entire sample.  With current data, we cannot be certain of how Arrokoth fits into the properties of the entire population.  Additional tight KBO systems are shown as well.  2011JY31, plotted with a green hexagon, was discovered to be a binary by New Horizons when it was close enough to resolve the system\cite{Weaver2022}.  An additional tight binary is (523764) 2014 WC510 that was discovered by an occultation observed by the RECON project\cite{Leiva2020}.  Clearly, there are tight binary and contact binary objects to be found.

Jupiter Trojans are interesting population to be compared to the KBOs.  The Lucy Mission is designed to investigate these objects at 5~AU.  As it turns out, three of the five prime targets are themselves binary objects.  All three binaries, Polymele, Eurybates, and Patroclus, small to large, are tight binaries.  Satellites of Eurybates and Patroclus were seen by direct imaging because of their much lower geocentric distance and better imaging scale.  These binaries would remain hidden from direct imaging detections were they in the Kuiper Belt.  Polymele, like Arrokoth, was discovered to be binary through occultation measurements.  The satellite of Polymele has an angular distance from the primary that prevents seeing it with direct imaging because of the compact size of the system.  Also of note is that the sizes (both relative and absolute) of the Polymele-system components are very similar to the two lobes of Arrokoth.

The purpose of Fig.~\ref{fig:bstat} is to show binary properties with a focus on the most compact systems as well as to show the parameter space where occultations are singularly effective.  The cluster of points at the top of the plot are beyond the range shown.  In the middle panel, 27 objects (53\%) are wider binaries with $a/R>100$.  It is interesting that the more loosely bound binaries are comprised entirely of smaller primaries and these are not well distributed with respect to the CCKBO size range shown and tend toward the smaller end of the range.  The number of CCKBOs at the smallest range does not look entirely consistent with the histogram in the top panel.  This is likely due to differing approaches in the two samples for estimating diameters and factor of two shift in sizes for either sample would eliminate the offset.  The cluster on the right side of the plot consist of 8 large objects, 16\% of the sample, all with $D>400$.  Two of these are CCKBOs.  It is also interesting that these large objects are all tight systems with $a/R$ in the range from 15-35 rather than a broader range extending to more loosely bound components.

The objects in the plot range of $D<400$ and $a/R<100$ are the focus of this paper.  There are 15 objects (29\%) from the sample that fall in this range.  All dynamical classes are represented here and in rough proportion to their fraction in the sample itself.  However, there is a lower bound in the sample at $D=100$ for tight CCKBOs.  Below this size, the only tight objects in the sample are Centaurs and these are easier to see compared to CCKBOs due to their smaller heliocentric distances.  From the direct detection sample, there appears to be a cutoff where no objects are seen for $a/R<25$ for $D<400$.  From this plot alone, it is hard to distinguish between an intrinsic population property or an observational bias.

The bottom panel of Fig.~\ref{fig:bstat} remaps the same sample of objects into an observational frame of reference.  Here, the minimum object distance in 2025 is used to show the best case angular separation for the binaries.  A separation of 40 mas corresponds to the size of a WFC3 pixel on HST\null.  The smallest angular separations seen in the sample are at $\sim$1.5-pixels.  With direct imaging, there is a clear observational bias making binaries harder to detect with decreasing angular separation and eventually impossible.  However, the three smallest separation objects were all found by other means, two are occultation discoveries and one is from New Horizons when it was sufficiently close.  Occultations excel in providing constraints for tight systems and equally effective with smaller primaries.  This method is also the best at uncovering contact binaries and flattened primordial shapes.  These data will be crucial for constraining the angular momentum budget during the formation process.

The SI model predicts formation of contact and tight binaries but the existing simulations do not tell us how often these binaries form, because this depends on unknown parameters (e.g., size of particles in the proto-planetary disk, metallicity, etc.), and what are their properties (e.g., the size ratio of the two components, shape parameters).  The next-level occultation observations will be game changing in that they will lead to detection of a statistically large number of contact and tight binaries. They offer an exciting possibility to compare the predictions of the SI model, and other planetesimal-formation models, with reality.  In so doing, we can then reverse-engineer the processes and conditions that contributed to the planetesimal formation in the Solar System and, by extension, elsewhere in exoplanetary systems. 

\section{Citizen Science Contributions}

With occultation campaigns, bringing more observation sites strategically into play almost always raises the quality and value of the science result. While a single chord on a distant object with high astrometric uncertainty provides critical data, acquiring this type of measurement reliably benefits from a coordinated network of telescopes to ensure that at least one positive detection. Collecting multiple chords on an object with low astrometric uncertainty can provide information on topography, cratering, and binarity; however, such results require intentionally positioning enough telescopes across the shadow path. The very nature of the research involved requires some level of coordination and distributed collection of data.  

While there are several important models of coordination involving both professional and amateur astronomers, more effort can be invested to further enhance the quality and impact of results accessible through occultations. Several collaborations involving professional astronomers have deployed resources provided through research and spacecraft mission funding on highly coordinated campaigns around the globe. In 2023, the International Occultation Timing Association (IOTA) had 348 participants contributing scientific results worldwide.  The stand-alone IOTA efforts are loosely coordinated activities.  On the other end of the spectrum, the Research and Education Collaborative Occultation Network (RECON) has established a transect of 40 telescope sites with a similar number of participants to IOTA from Arizona up into British Columbia with an emphasis on occultations of outer Solar System objects.  RECON is a very tightly organized effort.  There are other efforts that fall between these two extremes, such as the Lucky Star occultation program.  Through these efforts, the community has learned, shared, and grown in capacity and success in occultation work, but more can be done.

Coordinated occultation campaigns involve a diverse and motivating set of responsibilities and challenges for participants that can successfully be accomplished by novices and professional astronomers alike. Telescope and camera equipment designed for use by amateur astronomers can be effectively used for occultation research, which makes training novice and amateur astronomers possible for the successful collection of occultation data. Another important component of campaigns involves selecting, scouting, and practicing at deployment sites along an assigned occultation chord. This is most effectively and safely conducted by engaging a team of observers with diverse experiences and perspectives. Occultation campaigns inherently build communities and require collaboration and coordination for success in gathering data that must be geographically distributed across the shadow track. Recent campaign efforts for the Lucy mission showed that the collection of occultation data is highly accessible to individuals with all levels of background in astronomy.

Given these inherent elements of occultation research, engagement of large numbers of volunteer researchers on occultation campaigns provides a unique opportunity to contribute to the growing field of research in the realm of citizen science. The field of research into citizen science has grown and evolved over the past half century \cite{Vohland2021}.  Even the definition of citizen science is a complicated issue that includes discussions of science, education and even politics \cite{Haklay2021}.  Occultation research provides an exceptional platform for integrating volunteers into a community of researchers committed to advancing scientific knowledge about the Solar System, with accompanying benefits of enhancing their identity as researchers, building networks, and enhancing knowledge about scientific research within the general population.

Engagement of citizen science researchers in occultation campaigns provides an additional opportunity for investigation into participant learning \cite{Kloetzer2021}. Projects like RECON have focused on engagement of teachers and students in formal education settings, but it was found that some of the most successful engagements occurred in the transition space to after-school learning with astronomy clubs at several of these schools. It is also the case that individual researchers unaffiliated with schools and formal education have been significant contributors to occultation campaigns and have grown in astronomical observing skills and self-identification as researchers.

Occultation campaigns provide an excellent opportunity for participatory citizen science research. These campaigns bring groups of individuals together (either in person or remotely), provide training in skills related to astronomical observation, build compelling levels of community and collaboration, and motivate interest in STEM and STEM careers while also making valuable contributions to understanding solar system formation.  Occultation research requires distributed networks of observers and drawing from the ranks of citizen science volunteers has proven to be very effective on a global scale. Specifically, occultations of Arrokoth needed for the NASA New Horizons flyby and occultations of the NASA Lucy Mission Trojan asteroid targets have resulted in robust collaborations with researchers and citizen science volunteers in Senegal, Spain, Colombia, Mexico, Argentina, Japan, Korea, South Africa, Australia, and the United States. Through these efforts, citizen science volunteers were successfully trained to effectively and reliably collect quality occultation data of outer Solar System bodies.

\section{Conclusions}

Detailed shape determinations and the search for duplicity both imply large deployment efforts for occultation observations.  Direct detection of satellites requires covering a very large region of space round the primary.  Imaging searches are the best tools for more widely separated components.  Occultations are most effective at searching for very close companions.  So far, the most common explanation for distant satellites comes from collisions and this is the case across the entire solar system.  Tight systems ($a/R<20$) are much harder to explain with collisions and appear to be more directly tied to the initial formation process.  Even with these tight systems, the occultation coverage required is significantly larger than the body itself, requiring larger campaigns than those interested only in the primary.  A complementary constraint on satellites also comes from occultation astrometry that can place direct constraints on unseen masses when a sufficiently large spread of stations is impractical.  Shape determinations come from concentrating the stations on the main objects as much as possible.  Thus, the shape and duplicity goals drive occultation deployment plans in opposite directions.  The three-occultation scheme builds in a natural variation to cover both spread, for duplicity, and tight spacing, for shape.

With current levels and modes of funding, we might optimistically expect a few Arrokoth-class occultation campaigns per decade on other CCKBOs. The upcoming survey from the Rubin Observatory will significantly increase the number of well-observed KBOs and thus reveal many more opportunities for detailed occultation studies.  However, our ability to collect data is not limited by the number of occultation opportunities.  Instead, the real limit is the amount of support needed to collect a large enough sample.  A population study of CCKBOs fueled by detailed shape measurements requires a much larger coordinated effort than is feasible with current levels of effort.  Such a larger survey is now possible because of the results from the Gaia Mission and will be even better after just the first year of the Rubin survey.  Work remains to better quantify the size of survey needed and the requirements on the final data products.

A large coordinated occultation survey of CCKBOs requires a very unusual path forward compared to other large scientific investigations.  A typical discovery path is to build facilities, such as general purpose observatories (e.g., HST, JWST, TMT, VRO/LSST, Gaia) to tackle a big problem.  The VRO/LSST and Gaia projects are both in the \$1B project class, including operations and data analysis.  Another path is in situ investigation through NASA, ESA, JAXA, UAE Space Agency, and other space agencies to build a purpose-built machine to investigate a few objects from a nearby vantage point.  The survey concept presented here builds on these prior investments.  The one example given in section 2 is not the only choice or presumptive cost but already shows that a large occultation survey in a smaller class in term of total cost.  A compelling factor is its societal impact for the potential to involve more than an isolated science community.

The scientific promise of an occultation survey addresses fundamental questions in solar system formation that if they could be done with traditional methods, would be a strong candidate for the next great observatory or spacecraft.  Occultations require very basic and relatively simple equipment that almost becomes secondary to the cost and effort required.  Here, costs come entirely through the coordination and execution of the observational campaigns fueled by community scientists.  The advantage of this method is that we can collect a unique set of population statistics for the most ancient and undisturbed reservoir of small bodies.  This approach is also scalable to a size that can collect the large sample needed to drive the next generation of formation models.  It is now time to add this new and yet old tool to those of traditional observatories and space exploration missions.

\aucontribute{M. W. Buie -- occultations, J. M. Keller -- educational outreach, D. Nesvorn\'y -- formation models, S. B. Porter -- orbit fitting}
\ack{Thanks to the hundreds of observers that have contributed to occultation observations.}

\bibliographystyle{rsta}
\bibliography{references.bib}

\end{document}